# Revealing Method for the Intrusion Detection System

M.Sadiq Ali Khan

**Abstract**—The goal of an Intrusion Detection is inadequate to detect errors and unusual activity on a network or on the hosts belonging to a local network by monitoring network activity. Algorithms for building detection models are broadly classified into two categories, Misuse Detection and Anomaly Detection. The proposed approach should be taken into account, as the security system violations caused by both in-compliance with the security policy and attacks on the system resulting in the need to describe models. However, it is based on unified mathematical formalism which is provided for subsequent merger of the models. The above formalism in this paper presents a state machine describing the behavior of a system subject. The set of intrusion description models is used by the evaluation module and determines the likelihood of undesired actions the system is capable of detecting. The number of attacks which are not described by models determining the completeness of detection by the IDS linked to the ability of detecting security violations.

**Index Terms**—Intrusion Detection, Misuse Detection, Anomaly Detection,

——————————    ◆    ——————————

## 1 INTRODUCTION

Software still suffers from vulnerabilities that allow attackers to gain illicit access to computer systems. Attackers exploit vulnerabilities to hijack control of a process' execution as a means to access or alter a system as they desire. Intrusion detection system are software and/or hardware components that monitor computer systems and analyze events occurring in them for signs of intrusions. Due to widespread diversity and complexity of computer infrastructures, it is difficult to provide a completely secure computer system. Therefore, numerous security systems and intrusion detection systems that address different aspects of computer security. The design and construction of host-based intrusion detection systems is an active research area. Several papers in the intrusion detection have been published in the past [8],[9],[10]. However, the growth of the field has been very rapid, and many new ideas have emerged ever since these systems are invented.

## 1.1 Classification of attacks and intrusion

Several taxonomies that were developed later mainly focused on two issues: (i) categorization of computer misuse (i.e. attacks) and (ii) categorization of the people trying to get unauthorized access to computers (perpetrators), and the objectives and results of these attempts. Following are the common type of attacks:

————————————————

- *M.Sadiq Ali Khan is with the Department of Computer Science, University of Karachi, Karachi-Pakistan.*

1.1.1 **Denial of Service (DoS) attacks:** These attacks attempt to "shut down a network, computer, or process; or otherwise deny the use of resources or services to authorized users" [11]. An example of operating system attack is teardrop, in which an attacker exploits a vulnerability of the TCP/IP fragmentation re-assembly code that do not properly handle overlapping IP fragments by sending a series of overlapping packets that are fragmented. Typical example of networking DoS attack is a "SYN flood" attack. In this attack, attacker establishes a large number of "half-open" connections using IP spoofing. Other examples of DoS attacks include disrupting connections between machines thus preventing access to a service, preventing particular individuals from accessing a service, disrupting service to a specific system or person, etc. In distributed DoS (DDoS) attack, which is an advanced variation of DoS attack, multiple machines are deployed to attain this goal. DoS and DDoS attacks have posed an increasing threat to IDS and techniques to thwart them have become an active research area [12],[17],[18],[19].

1.1.2 **Probing attacks:** These attacks scan the networks to identify valid IP addresses and to collect information about them (e.g. what services they offer, operating system used). These attacks are probably the most common ones, and are usually precursor to other attacks. Examples of probing attacks include IPsweep (scanning the network computers for a service on a specific port of interest), portsweep (scanning through many ports to determine which services are supported on a single host), nmap (tool for network mapping), etc.

**1.1.3  R2L (Remote to Local) attacks:** Where an attacker who has the ability to send packets to a machine over a network, gains access to the machine. In most R2L attacks, the attacker breaks into the computer system via the Internet. Typical examples of R2L attacks include guessing passwords (e.g. guest and dictionary attacks) and gaining access to computers by exploiting software vulnerability.

**1.1.4  U2R ( User to root ) attacks:** Where an attacker who has an account on a computer system is able to misuse/elevate her or his privileges by exploiting a vulnerability in computer mechanisms, a bug in the operating system or in a program that is installed on the system. Unlike R2L attacks, where the hacker breaks into the system from the outside, in U2R compromise, the local user attacker is already in the system and typically becomes a root or a user with higher privileges. The most common U2R attack is buffer overflow.

## 1.2  Host-Based IDS

Host based intrusion detection systems analyze users activities and behavior on a given machine. However, depending upon the processing performed; host-based IDSs can significantly impact the performance of the machine they are running on. In addition, audit sources used in host-based intrusion analysis, can be easily modified by a successful attack, which represents another limitation of host-based IDSs. In order to eradicate these drawbacks, host-based IDSs have to process the audit trail sufficiently fast to be able to raise alarms before an attacker has an opportunity to observe and/or modify the audit trail or the intrusion-detection system itself. There are several types of information that are typically used in host-based IDSs, e.g. (i) system commands, (ii) system accounting, (iii) syslog and (iv) security audit information. Automated detectors find attacks without human interaction. The goal of automated detection is to maximize the number of actual attacks discovered while minimizing the number of false alarms.

**1.2.1  System commands:** System commands are a useful source of information that can be employed by host based IDSs for detecting malicious users [20],[21]. By analyzing system commands that users invoke in their sessions, it is possible to build user profiles, which describe users' characteristics and common behavior.

**1.2.2  System Accounting:** System accounting is present in both Windows and Unix operating systems. Although the interest for system accounting in Windows environment is increasing, there have not been many intrusion detection approaches that used this type of data for intrusion analysis [7].

**1.2.3  System log information:** System log data contains information that is not available at the network level, such as when users log in, when they send email, who they send email to, which ftp logs commands are issued, and which files are transferred. One of the major drawbacks of using syslog information for intrusion detection is that syslog information is not very secure, since several syslog daemons exhibit buffer overflow exploitation [6].

**1.2.4  Security audit processing:** The security audit trails represent records that contain all potentially important activities related to the security of the system. In addition, advantages of using security audit data include strong user authentication, easier audit system configuration, and fine-grain parameterization of collected information [4]. Several research groups [2],[3] have been actively using security audit trails mainly for host-based intrusion detection systems. The focus of their research has been mainly to define what information the security audit trail should contain in order to increase the IDS prediction performance as well as to establish an acceptable common format for audit trail records.

## 1.3  Misuse vs Anomaly Detection

Many contemporary IDSs integrate both approaches to benefit from their respective advantages [5],[16].

**1.3.1  Misuse Detection:** Misuse detection is the most common approach used in the current generation of commercial intrusion detection systems (IDSs).

- **Signature-based techniques:** In signature-based IDSs, monitored events are matched against a database of attack signatures to detect intrusions. In addition, once a new attack is discovered and its signature is developed, often there is a substantial latency in its deployment across networks [1].

- **Rule-based systems:** Rule-based systems use a set of "if-then" implication rules to characterize computer attacks. In rule-based IDSs, security events are usually monitored and then converted into the facts and rules that are later used by an inference engine to draw conclusions.

- **State transition analysis:** Intrusion detection using state transition analysis requires the construction of a finite state machine, in which states correspond to different IDS states, and transitions characterize certain events that cause IDS states to change. Every time when the automation reaches a state that is flagged as a security threat, the intrusion is reported as a sign of malicious attacker activity.

**1.3.2  Anomaly Detection:** Increase in the number of computer attacks, in their severity and complexity has raised substantial interest in anomaly detection algorithms due to their potential for recognizing unforeseen and emerging cyber activities.

## 2. IDS Architecture

It is very important that the security mechanisms of a system are designed so as to prevent unauthorized access to system resources and data. Regardless of the type of IDS there are a few common components that typically constitute an IDS:

**2.1 Traffic Collector:** The component is responsible for gathering activity and event data for analysis. On a host-based IDS this will typically include metrics such as inbound and outbound traffic and activity recorded by the operating system in log and audit files.[5]

**2.2 Analysis Engine:** The analysis engine is responsible for analyzing the data gathered by the traffic collector. In case of a knowledge-based IDS the data will compared against a signature database.

**2.3 Signature Database:** Used in knowledge-based systems, the signature database contains a collection of signatures known to be associated with suspicious and malicious activities. It could be said that a knowledge based IDS is only as good as its database.

**2.4 Management Interface:** A management interface providing a mechanism by which system administrators may manage the system and receive alerts when intrusions are detected.

A host-based IDS runs directly on a server or desktop system and uses the resources of that system to examine log and audit files together with network traffic entering and leaving the system. A false positive is legitimate and authorized activity on a system which is incorrectly identified by an IDS as being suspicious or malicious. By running directly on the host and analyzing log files in context with overall system activity the number of false positives is reduced.

## 3. Development of the Analysis Module

This article offers the description of an approach to status security analysis aimed at detecting information security violations in the course of computer system operation. The online system condition monitoring subsystem should detect system transfers to unsafe condition due to intrusions into the system. Hence, the task of online system status security monitoring consists in:
*i) Detecting the conditions that contradict to the security policy determined in the system*
*ii) Identifying the reasons that caused an insecure condition of the system;*
*iii) Evaluating the security of the system being intruded.*

The approach proposed should take into account security system violations caused by both incompliance with the security policy and attacks on the system resulting in the need to describe models[15].

### 3.1 Identification of System Model for Security breach

Let us introduce definitions of the concepts to be used in further reasoning. $\{S_b\}$–a huge amount of subjects ;$\{O_{bj}\}$–a huge amount of objects; $\{Opt\}$ – a massive amount of operations; $\{Pgs\}$ – a massive amount of services used by the programs or program interfaces. The introduction of this huge amount into the system model description is due to the fact that system subject operations over objects are implemented using services. Hence the matrix of subject to object access in the system may be defined as follows: $Mc'(S_b, Pgs, O_{bj})$ – access matrix for programs used on behalf of subjects to perform operations with system objects.

Then machine $A = \{\sigma, t, Output, \sigma_0, \delta, \lambda\}$ which represents the user performance with respect to the security policy determined in the system, may be described as follows: $\sigma = \{Opt_1(prog_1, 0_1), \ldots, Opt_j(prog_j, 0_j)\}$ – a machine state describing operations performed by a system subject over objects; the huge amount of condition is partially rank-ordered. $t \in Opt(Pgs_j, o_k)$ – controlling machine symbols corresponding to the operations performed by the subject over system objects using programs. A secure status is a condition describing operations performed by the subject that do not violate to the security policy. Thus the condition security evaluation describes the machine exit as Output = {Secure, UnSecure}. The machine completes its operation if it goes to an unsafe condition. Then the transition function $\delta$ may be described as follows:

$$\forall t = Opt_j \, \exists \, Opt_j(prog_j, 0_j) \in \sigma_i = t \twoheadrightarrow \sigma_i$$
$$\forall t = Opt_j \, !\exists \, Opt_j(prog_j, 0_j) \in \sigma_j = t \twoheadrightarrow \sigma_j \cup t$$

The exit function of machine $\lambda$ may be presented as follows:
$\forall t = Opt_j \, (Opt_j(prog_j, 0_j) \in \sigma_j) \, V(Opt_j(prog_j, 0_j) \in Mc'(S_b, Pgs, o)) \twoheadrightarrow Output = Secure$
$\forall t = Opt \, (Opt_j(prog_j, 0_j) \in \sigma_j) \, V(Opt_j(prog_j, 0_j) \in Mc'(S_b, Pgs, o)) \twoheadrightarrow Output = UnSecure$

### 3.2 Detection Model

Let us discuss a model describing likely system attacks. System attacks are identified with the use of attack signatures $\{Signt_n\}_{n=1}^{M}$

The massive amount of signatures describing attacks may be grouped into submultitudes according to their PROPERTY. The PROPERTY reflects the huge amount of attack signatures into multitude $prpty_m \, m \in 1: N$ that describes the attack objectives. Each element of multitude $prpty_m$ reflects the objectives of an attack involving signatures. Multitude $\{prpty_m\}_{m=1}^{N}$ is partially rank-ordered. It is important that the intruder performing intrusion advances in its actions by means of launching various attacks on the system. Then the huge amount of signatures may be rank-ordered in accordance with the intrusion stage as:

$\{Signt_{j1}\}^{M1}$ ...... $\{Signt_{j1}\}^{Mk}$ while $\sum_{i=1}^{k} m_i = M$
$j_1=1$ $\quad\quad\quad\quad j_k=1$

At that, the scenarios of security violation (intrusion) may be described as Scn=(Signt$_0$, Signt$_1$, ..., Signt$_k$) k <= M provided that:
1) $\forall j, k \in 1: m, j \neq k \rightarrow Signt_j \neq Signt_k$
2) $\forall j, k \in 1: m, j <= k \rightarrow prpty(Signt_j) \leq prpty(Signt_k)$

The machine describing system security violations may be introduced using the following definitions: $\sigma = \{Signt_n\}$ – machine condition described by a signature corresponding to the most advanced intrusion phase reached by the intruder.
t $\in Signt_j$ – machine control symbols; prpty($\sigma_j$) for the current machine condition – machine exit
$\sigma_0 \in \sigma$ – initial condition in which the subject starts interacting with the system.

Machine transition function $\delta$ may be described as follows:
$\forall t = Signt_k \exists Signt_j \in \sigma_j: prpty(Signt_k) \leq prpty(Signt_j) \rightarrow \sigma_{j+1} = \sigma_j$

Output function $\lambda$ is described as follows:
$\forall t = Signt_j$ Output = prpty($\delta(\sigma_i, t)$).

## 3.3 Unified Model

The components describing the system model may be described using a unified structure. In the above structure machine condition will be described as follows: $\sigma = \{\{Opt_1(prog_1,0_1),....,Opt_i(prog_i,0_i)\}, Signt_n(Scn_l)\}$.

The entry of the unified machine is a user-performed operation using a service, or a user-performed system attack using a service. Hence, t $\in (Opt(Pgs_j, o_k) \vee Signt_n)$ – machine control symbols. Transfer function $\delta$ of the unified machine is described as follows:

$\forall t = Signt_k \exists Signt_j \in \sigma_j: prpty(Signt_k) <= prpty(Signt_j) \rightarrow \sigma_{j+1} = \sigma_j$

$\forall t = Signt_k !\exists Signt_j \in \sigma_j: prpty(Signt_k) <= prpty(Signt_j) \rightarrow \sigma_{j+1} = \sigma_j \cup Signt_j$

$\forall t = Opt_j \exists Opt_j(prog_j, 0_j) \in \sigma_i = t \rightarrow \sigma_{i+1} = \sigma_i$
$\forall t = Opt_j !\exists Opt_j(prog_j, 0_j) \in \sigma_j = t \rightarrow \sigma_{j+1} = \sigma_j \cup t$

The machine exit is the unified machine condition profile. In accordance with the above definitions the condition may be: safe, unsafe and attack condition. The exit function of the unified machine $\lambda$ is described as follows:

$\forall t = Signt_j$ Output=prpty($\delta(\sigma_i, t)$)

$\forall t = Opt_j (Opt_j(prog_j, 0_j) \in \sigma_j) \vee (Opt_j(prog_j, 0_j) \in Mc'(S_b, Pgs, o)) \rightarrow$ Output = Secure

$\forall t = Opt_j (Opt_j(prog_j,0_j) \in \sigma_j) \vee (Opt_j(prog_j,0_j)$ not member of Mc' (S$_b$, Pgs,o)) $\rightarrow$ Output = UnSecure

Therefore, this article offers a machine model providing for online monitoring of system condition security. Based on online monitoring of system condition security various security violations of the computer system may be detected.

## 4. Development features of the Acquisition Module

The attacks and intrusions themselves are commonly described as in lower level terms. The bridging of this gap should be facilitated by an adequate data acquisition method with an option to transform the data obtained to higher presentation levels. Even though for data acquisition in host-based IDS it is possible to use standard tools of operating system audits it is advisable to customize the data acquisition modules due to the fact that standard audit tools frequently acquire information useless for detecting system security violations while vital information is often missing. The level selection will be determined by two contradictory factors – the ease of information acquisition and the unambiguous decision-making process. Function $Level(Mk) = \{LevMod_m\}$ provides for the return of the level huge amount of operation and object descriptions used to describe intrusion model j. Let us assume that a acquisition module gathers data at level LevData. Let us designate the presentation of system objects and operations at the level of the acquisition module as Op$_{LevData}$, Pr$_{LevData}$, and at the model level as Op$_{LevMod}$, Pr$_{LevMod}$. Thereafter the proposed system will make it possible to describe the basic properties of the IDS determined by the data acquisition module.

1) **Validity of the intrusion detection system.** The data acquisition module should run at a minimum level of operation and object presentation present in a multitude of models describing attacks $\forall Mk$ LevData $\leq min(min(LevMod(Mk)))$. In the event that the above condition is not met the data acquisition module will not be capable of transferring complete information to the system event analysis module, and the operation of the IDS will be invalid.

2) **Compatibility of the modules of the IDS.** The data obtained by the data acquisition module of the host-based IDS should be reduced to a single format used by the analysis module of the intrusion detection system $\forall M_k \exists! F, G: Op_{LevMod} = F(Op_{LevData}), Pr_{LevData} = G(Pr_{LevData})$. The existence of single transformations F and G and their complexity determine the possibility to identify objects and operations at the level of intrusion model presentation as well as the complexity of development.

3) **Compatibility of the IDS with the computer system.** Ideally, the intrusion detection system should be transparent for the user. However, a module developed after such a pattern may prove inefficient because the volume of data acquired for analysis may be redundant. The use of data acquired on the events in a particular subsystem may be appropriate from the perspective of detecting subsystem attacks which limits system. Thus, a hit-hit option would provide for tracking the subsystem events that are critical for the system security. In accordance with the requirements the modification of the software environment may impact the compatibility of the data acquisition module with the computer system. As a result, it is recommended to use the intrusion detection system without modifying the software environment or using modification of environment variables. As shown above, the choice of performance level for the data acquisition module, in its turn, has an impact on the validity of IDS and the compatibility of its modules. At that, it is most appropriate to develop modules for acquisition of the data related to the intercept of system calls and API requests. Table 1 shows a comparison of the most common methods used to build a data acquisition module for Windows OS family host-based IDS. The number of models used and reduces the wholeness of the intrusion detection. It should be noted that the values shown in the above table for different relative features with respect to the method used for building a data acquirement module for host-based IDS, will be typical of not only Windows OS but of other operating systems too. The most promising methods for building a data acquisition component should be the method based on the intercept of system calls, Performance tools and the method based on the use of OS drivers. The use of methods based on the intercept of system calls is less lengthy compared to the method based on the use of OS drivers. However the use of methods based on the intercept of system calls requires strong efforts aiming at ensuring protectibility compared to the methods based on the use of OS drivers. As it follows from Table 1, the methods based on the intercept of system calls may also suffer from problems related to the requirement regarding protectibility. The methods based on the performance tools may also have deficiencies regarding the transparency of the user and the analysis of the system operation.

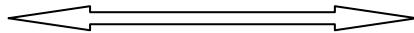
**Comparison Criteria**

| Methods | Standard Audit | System Calls | OS drivers | Shells | Performance Tools |
|---|---|---|---|---|---|
| **High degree of information contents** | ⇓ | ⇑ | ⇓ | ⇑⇓ | ⇑⇓ |
| **Ability to acquire info** | ⇑ | ⇑⇓ | ⇓ | ⇑ | ⇑ |
| **Methods of attack detection** | ⇑ | ⇑ | ⇑ | ⇑ | ⇑ |
| **Anomaly detection methods** | ⇑⇓ | ⇑ | ⇑ | ⇑ | ⇑ |
| **Analysis of system operation** | ⇑⇓ | ⇑ | ⇑ | ⇑ | ⇓ |
| **Versatility** | ⇑ | ⇑ | ⇓ | ⇓ | ⇓ |
| **Transparency for user** | ⇑ | ⇑ | ⇑ | ⇓ | ⇓ |
| **User-free info acquisition** | ⇑ | ⇑ | ⇑ | ⇓ | ⇓ |
| **Protectibilty** | ⇓ | ⇓ | ⇑ | ⇓ | ⇓ |

Table: Methods used to build a Data Acquisition Model

## 5. Conclusion

Thus, in this paper we've tried to find out the balance between intellectuality and usability of host-based IDS. Our algorithm of detection is efficient and communicative and can be used in practical IDS with different perspective. As OS drivers have more efficiency with less audit feature for attack detection and less protectibility for system calls.Whilst host-based intrusion detection systems work well for deployment on smaller numbers of systems i.e the tracking, monitoring and maintaining of hundreds or thousands of systems can quickly become a cumbersome overhead in terms of costs and resources. Proposed model makes it possible to justify the selection of the method for building a data acquisition module of host-based IDS.